\documentclass[apjl]{emulateapj}
\usepackage{epstopdf}
\usepackage{natbib}
\usepackage{amsmath}
\usepackage{enumerate}
\begin{document}
\title{Detectability of Local Group Dwarf Galaxy Analogues at High Redshifts}
\author{Anna Patej\altaffilmark{1} and Abraham Loeb\altaffilmark{2}}
\altaffiltext{1}{Department of Physics, Harvard University, 17 Oxford St., Cambridge, MA 02138}
\altaffiltext{2}{Department of Astronomy, Harvard University, 60 Garden St., Cambridge, MA 02138}

\email{E-mail: apatej@physics.harvard.edu, aloeb@cfa.harvard.edu}

\begin{abstract}
The dwarf galaxies of the Local Group are believed to be similar to the most abundant galaxies during the epoch of reionization ($z\gtrsim6$). As a result of their proximity, there is a wealth of information that can be obtained about these galaxies; however, due to their low surface brightnesses, detecting their progenitors at high redshifts is challenging. We compare the physical properties of these dwarf galaxies to those of galaxies detected at high redshifts using \textit{Hubble Space Telescope} and \textit{Spitzer} observations and consider the promise of the upcoming \textit{James Webb Space Telescope} on the prospects for detecting high redshift analogues of these galaxies.
\end{abstract}

\section{Introduction}\label{s:intro}
Local Group dwarf galaxies allow investigation into a wide range of astrophysical and cosmological processes; in addition to being representative of the most populous type of galaxy in the universe, their nearness enables detailed investigations of their stellar populations \citep[for a recent overview, see][]{mcconnachie12}. Due to their low masses and typically old stellar populations, many of these dwarfs are believed to have had the majority of their stars produced at early cosmic times and then had further star formation suppressed by reionization at redshifts $z\sim6-10$ \citep[e.g,][]{bullock00, ricotti05,loeb13}. Observational arguments in favor of this interpretation for some of the dwarfs have been based on their statistical properties \citep{bovill09} and star formation histories \citep{weisz14,weisz14b}. In this scenario, some of the present-day dwarfs should be similar to their progenitors at higher redshifts.

The advent of optical and infrared space-based telescopes -- the \textit{Hubble Space Telescope} (\textit{HST}) and the \textit{Spitzer Space Telescope} -- has allowed for the identification of numerous high-redshift ($z\gtrsim6$) galaxies, whose properties, including their sizes, star formation rates, and masses, have now been examined in detail \citep[e.g.,][]{stark09,labbe10,oesch10,bouwens11,ellis13,ono13}. However, current observations are missing a population of fainter galaxies that are needed to reionize the universe at these high redshifts \citep[e.g.,][]{alvarez12,finkelstein12,bouwens15,robertson15}. Discovering some of these fainter galaxies will be within the purview of future observatories like the \textit{James Webb Space Telescope} (\textit{JWST}).

Part of this population of fainter galaxies is likely to consist of the progenitors of galaxies like the Local Group dwarfs. \citet{boylan15} used an analysis of the UV luminosities of the dwarfs to determine that \textit{JWST} should be able to detect progenitors of galaxies like the Large Magellanic Cloud. Here, we compare the physical properties of the local dwarfs and the high-redshift galaxies that have already been detected, and place them in the context of the predicted detection limits for \textit{JWST} to examine the fraction of dwarf progenitors -- and thus the fraction of missing light -- that may be observable in the near future. Throughout our discussion, we use the standard cosmological parameters $\Omega_m=0.3$, $\Omega_{\Lambda}=0.7$, and $H_0=70$ km/s/Mpc. 

\section{Data}\label{s:data}
We obtain data for over 100 local galaxies from \citet{mcconnachie12}, including their $V$ band Vega magnitude $m_V$, half-light radius $r$, ellipticity $\epsilon$, and average metallicity $\langle \left[\mathrm{Fe/H}\right] \rangle$. To provide direct comparison with the high-redshift data, we convert $r_h$ to the circularized half-light radius, $r_h = r\sqrt{1-\epsilon}$, that is commonly employed. We select only the 87 galaxies that have all these quantities measured in \citet{mcconnachie12}, and note in particular that this excludes the Large and Small Magellanic Clouds (LMC and SMC). 

We use metallicity as an input to the Flexible Stellar Population Synthesis (FSPS) code \citep{conroy09,conroy10} to scale the galaxies back to $z=6,7$. Following the results of \citet{weisz14}, who analyzed the star formation histories (SFHs) of a subsample of 38 of these dwarfs, we remove those galaxies whose SFHs indicate that the majority of their stellar populations were formed later than these redshifts. We keep those whose SFHs are consistent with at least $50\%$ (within errors) of the stars having been formed prior to $z=6,7$, as well as all the remaining galaxies from \citet{mcconnachie12} whose SFHs have not yet been measured, for a total of 73 galaxies. We use a delayed tau-model with $\tau = 0.2$~Gyr, which assumes an early starburst such that nearly all the stars we see today already existed at the redshifts of interest. The code also calculates the evolution to $z=0$, from which we take the predicted $V$ band magnitude and compare it to $m_V$ from \citet{mcconnachie12} to obtain a correction for the stellar mass of each galaxy; we then use these values to correct the $z=6,\:7$ magnitudes since we assume that all the stars we see today were already present at those high redshifts.

The corresponding parameters for observed $z=6$ and $z=7$ galaxies are obtained from several sources. The \textit{Spitzer} IRAC 3.6 $\mu$m fluxes of a sample of $z\sim6$ galaxies are based on \citet{gonzalez12}, and the 4.5 $\mu$m fluxes of $z\sim7$ galaxies are taken from \citet{labbe10}. We adopt the $2\sigma$ lower and upper limits on the fluxes, corrected to S/N = 5, which gives $F_{3.6} = [38.9, \:667.0]$ nJy and  $F_{4.5} = [13.3,\:637.5]$ nJy. \cite{oesch10} provides a value of $r_h = 0.7\pm0.3$ kpc for detected galaxies at these redshifts, measured from near-infrared observations using \textit{HST}. We again adopt the $2\sigma$ bounds on this quantity; accordingly, the range of observed sizes that we use is $0.1-1.3$ kpc. 

The future \textit{JWST} mission's NIRCam imager will have two filters, F356W and F444W, which will cover the bandwidth of the \textit{Spitzer} filters above. For the purposes of this comparison, we will assume that the \textit{Spitzer} 3.6 $\mu$m filter is identical to F356W and the 4.5 $\mu$m filter is identical to F444W.

\section{Comparison of Physical Parameters}\label{s:analysis}
\begin{figure*}
\begin{center}
\includegraphics[scale=0.49]{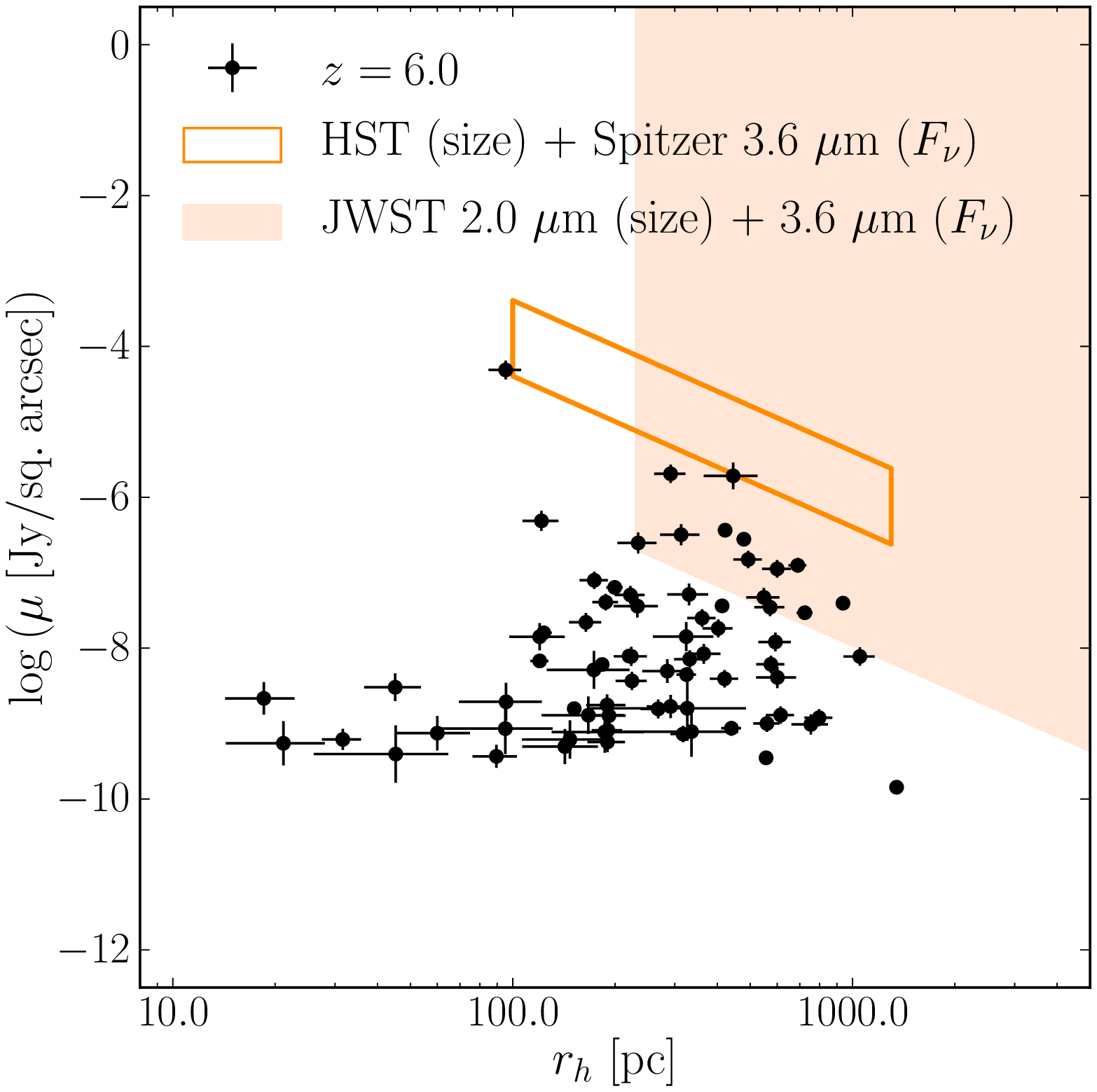}
\includegraphics[scale=0.49]{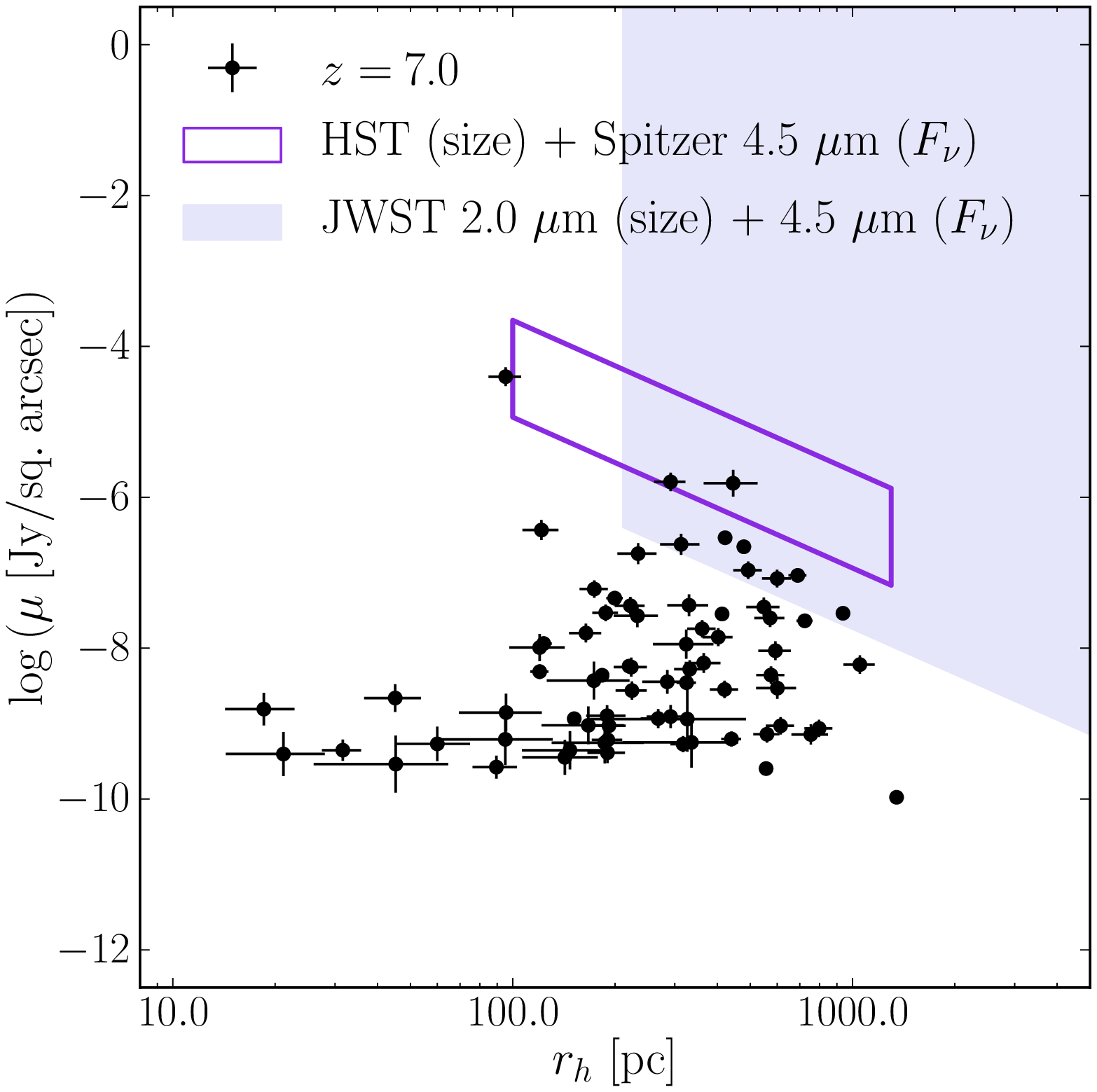}
\end{center}
\caption{The surface brightness-size relation for $z\sim6$ (left) and $z\sim7$ (right) galaxies using \textit{Spitzer} 3.6 $\mu$m and 4.5 $\mu$m fluxes, respectively, as well as sizes measured from \textit{HST}. The region bounded by solid lines indicates the portion of the parameter space that has been observed by current programs. The shaded region indicates the part of the parameter space that should be accessible to \textit{JWST}, with the size limit coming the diffraction limited radius in the given filter (see Section~\ref{s:jwst}).}\label{f:size_sb}
\end{figure*}

We use the FSPS-derived fluxes and the sizes $r_h$ to calculate the surface brightness in Jy/$\mathrm{arcsec}^2$ for the dwarfs at $z=6$ and $z=7$. We selected those galaxies with cumulative SFHs $\gtrsim0.5$ (within the error profile) at early times from \citet{weisz14} and used $\tau=0.2$ Gyr to scale them, as discussed in Section~\ref{s:data}, as well as all the remaining galaxies that do not have measured SFHs. We additionally plot the region bounded by the $2\sigma$ limits on the surface brightness and half-light radius for galaxies detected at $z\sim6$ and $z\sim7$, using \textit{Spitzer} 3.6 $\mu$m and 4.5 $\mu$m fluxes, respectively. Figure~\ref{f:size_sb} shows these regions alongside the scaled galaxies selected from \citet{weisz14} and the remaining galaxies from \citet{mcconnachie12}. From this, we see that virtually none of the local dwarf analogues have been detected yet.

Nevertheless, the sizes of the dwarfs and the high redshift galaxies agree extremely well, excluding the extreme smallest and faintest objects. However, we do not include in our comparison a scaling of $r_h$ with redshift. Luminous galaxies at higher redshifts have been observed to have their sizes scaled by a factor of $(1+z)^{-s}$, where $s$ is in the range of $1.0-1.5$ \citep[e.g.,][]{oesch10,mosleh12}. There is some indication, however, that at the lowest luminosities yet studied, the half-light radius remains approximately constant with redshift \citep[see Figure 12 of][]{ono13}. Our results are consistent with the notion that the Local Group galaxies had roughly the same size at high redshifts as they have at present.

\section{Predictions for \textit{JWST}}\label{s:jwst}
\textit{JWST} will rely on the NIRCam imager \citep{rieke05,beichman12} to obtain photometry of high-redshift galaxies. We use the prototype exposure time calculator (ETC) for NIRCam\footnote{http://jwstetc.stsci.edu/etc/input/nircam/imaging/} to compute the signal-to-noise ratio for the dwarf galaxies scaled back to high redshifts. We assume a total of 100 hours of exposure time; with such a set-up, the ETC predicts that a point source flux of 1.0 nJy can be detected in F356W and 2.0 nJy in F444W. 

We can then calculate the minimum surface brightness for the local galaxy analogues in each band. We assume that, as is currently done with \textit{HST}/\textit{Spitzer} observations of high redshift galaxies, the size of the galaxies is measured from bands at shorter wavelengths. Accordingly, in Figure~\ref{f:size_sb}, we also plot a shaded region bounded by the diffraction-limited radius in the 2.0 $\mu$m filter at these redshifts and by the surface brightness corresponding to the S/N=5 fluxes calculated by the ETC. We caution, however, that when comparing the sizes care must be taken, as the data for the galaxies uses the half-light radius, whereas the minimum size prediction for \textit{JWST} is given by the radius of a high redshift object in the diffraction limit. 

From Figure~\ref{f:size_sb}, we can see that \textit{JWST} can be expected to discover some of the local dwarfs if their stars already formed at early cosmic times. In particular, we predict that roughly 60-65\% of the combined light of the dwarfs will be accessible to \textit{JWST}. This corresponds to a detection of 9/73 dwarfs at  $4.5\;\mu$m and 13/73 dwarfs at $3.6\;\mu$m, respectively. This differs from the result of \citet{boylan15} primarily due to our uniform assumption that these galaxies formed most of their stars at very early times. If it is the case that significant star formation occurs late in the galaxy's evolution, then there will not be enough light emitted to render the galaxy detectable even by \textit{JWST}. Accordingly, these predicted fluxes will need to be scaled by the fraction of stars formed by $z\sim6,7$ once more of the galaxies have their star formation histories analyzed.

\begin{figure*}
\begin{center}
\includegraphics[scale=0.49]{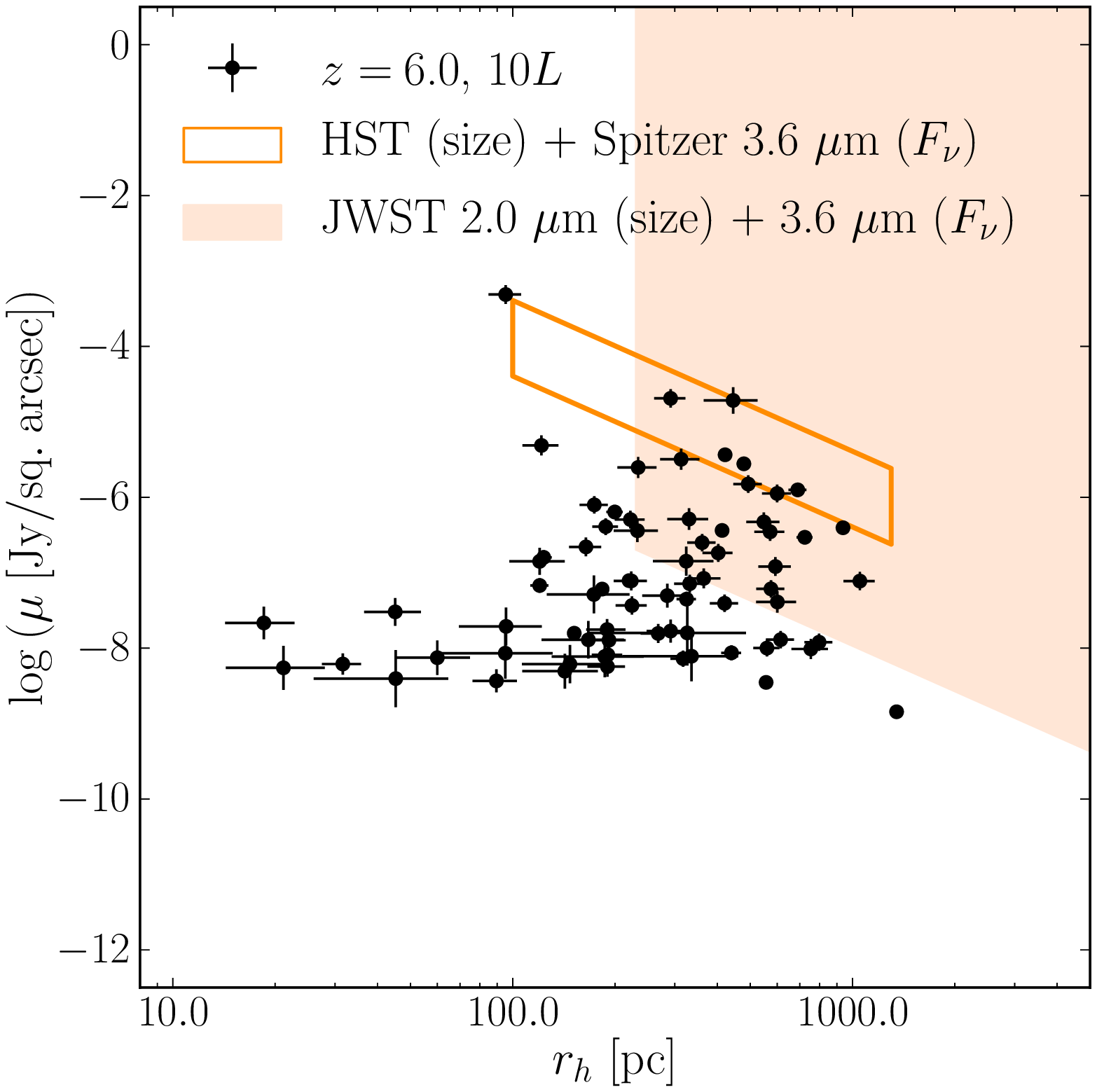}
\includegraphics[scale=0.49]{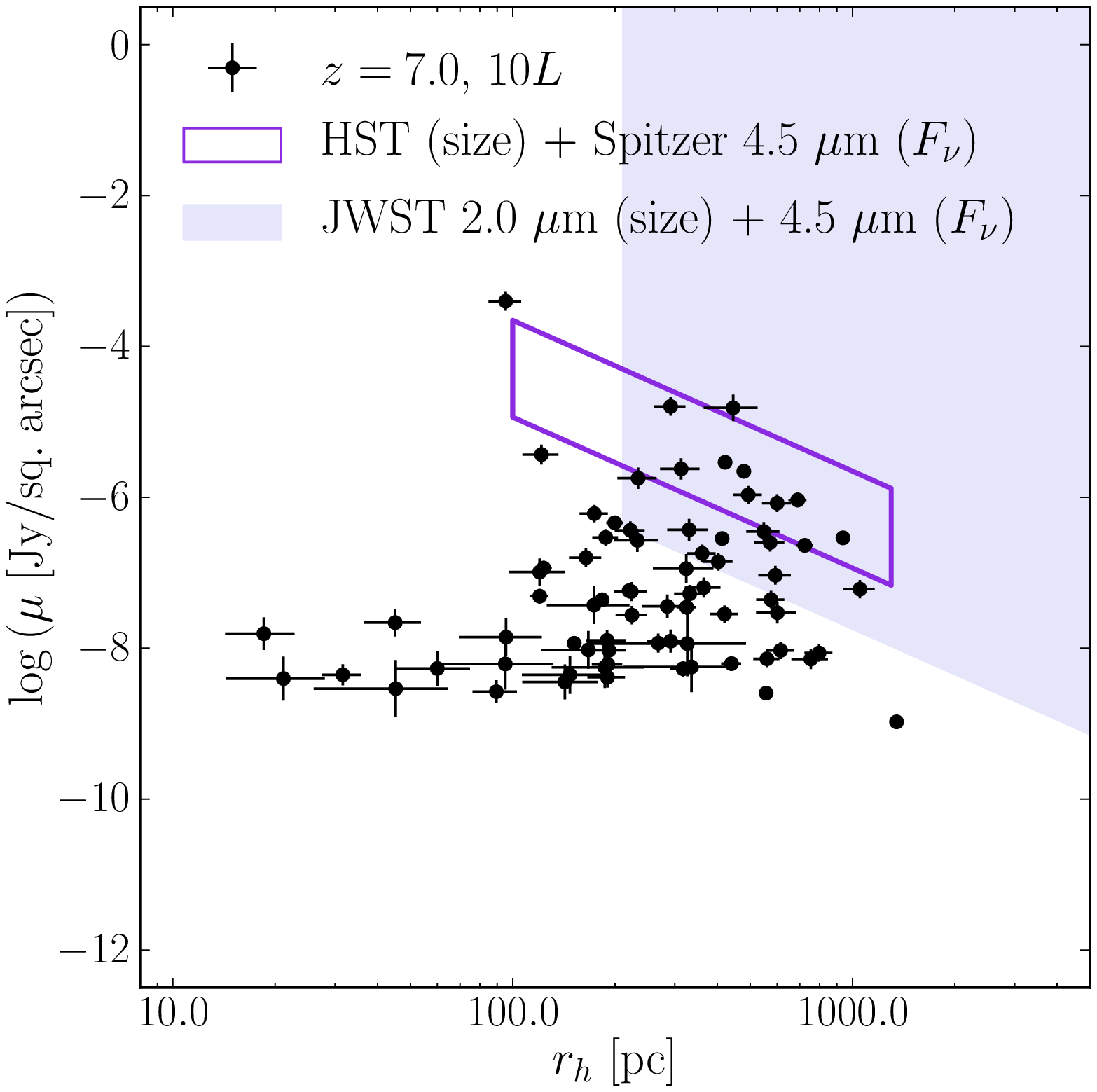}
\end{center}
\caption{Same as Figure~\ref{f:size_sb}, but assuming the ancient stars were 10 times more luminous than those in the dwarfs today.}\label{f:size_sb_10}
\end{figure*}

Our calculated fluxes are predicated upon the assumption that the stellar populations of the $z=0$ dwarfs are the modern analogues of the stars at $z\sim6-7$. However, these stars may be supplemented by PopIII-like stars, which are predicted to have masses in the range $10-1000M_{\odot}$ and short lifetimes \citep{abel02,prescott09,cassata13,loeb13,sobral15}. Accordingly, in Figure~\ref{f:size_sb_10} we consider the scenario in which the dwarf galaxies are 10 times more luminous at high redshifts due to an ancient population of stars that no longer exists.

In this case, we find that some of the dwarf progenitors could be among the galaxies already found with \textit{HST} and \textit{Spitzer}. \textit{JWST} would observe $\sim67\%$ of the light of the Local Group dwarfs. This scenario can be distinguished from the fiducial one based on spectroscopy; several theoretical works have indicated that such a massive stars would have strong helium line emission, which would distinguish these stellar populations \citep{bromm01,tumlinson01,schaerer03}. 

An alternative source of an increase in surface brightness at early times may be provided by tidal stripping. The simulations of \citet{penarrubia08} illustrate the effect of tidal stripping on dwarf galaxies by massive halos. This effect decreases both the sizes and surface brightnesses of the dwarf galaxies. If tidal stripping has played a major role in the history of the observed Local Group, then it is likely that at higher redshifts they were both brighter and larger, which would make the analogues of these galaxies more likely to be detectable. 

In addition to the enhanced capabilities of \textit{JWST} relative to \textit{HST} and \textit{Spitzer}, we note that further gains in sensitivity can be made using gravitational lensing \citep{mashian13,atek15}. Lensing has already permitted the discovery of high redshift galaxies even smaller than the sizes considered here \citep[e.g.,][]{kawamata15}; this technique using \textit{JWST} is likely to be able to probe a larger sample of dwarf progenitors.

\section{Conclusions}
We have compared the physical properties of Local Group dwarf galaxies to high-redshift galaxies. We find that the sizes of the two populations agree very well, but when translated to higher redshifts, these dwarfs are too faint to be detected at present. However, in a deep field, the upcoming \textit{JWST} mission should be able to detect analogues of the brightest of these objects, corresponding to about $60\%$ of the total light of the dwarf population (omitting the LMC and SMC, and galaxies without measurements), assuming that their stars formed early. This fraction increases if we assume a population of ancient, massive stars in these galaxies at high redshifts; spectroscopy and number counts will enable us to distinguish these two scenarios. Additionally, if these dwarfs have been significantly affected by tidal stripping, then this effect can also amplify the potential for analogues of these galaxies to be detected at high redshifts. 

\acknowledgments

We would like to thank Charlie Conroy and Matt Walker for helpful comments on a draft of this paper. We are also grateful to Charlie Conroy for useful discussions regarding the FSPS code.

This work was supported by the National Science Foundation Graduate Research Fellowship under Grant No. DGE-1144152 and by NSF Grant No. AST-1312034.

This analysis made use of Numpy \citep{numpy11} and Matplotlib \citep{hunter07}.

\end{document}